\documentclass[runningheads,a4paper]{llncs}

\usepackage{amssymb}
\usepackage{graphicx}

\usepackage{amsmath}
\usepackage{stfloats}
\usepackage{amsfonts}

\usepackage{algorithm,algorithmic}

\usepackage{url}
\urldef{\mailsa}\path|{alfred.hofmann, ursula.barth, ingrid.haas, frank.holzwarth,|
\urldef{\mailsb}\path|anna.kramer, leonie.kunz, christine.reiss, nicole.sator,|
\urldef{\mailsc}\path|erika.siebert-cole, peter.strasser, lncs}@springer.com|
\newcommand{\keywords}[1]{\par\addvspace\baselineskip
\noindent\keywordname\enspace\ignorespaces#1}

\begin{document}

\mainmatter  

\title{Feature Bagging for Steganographer Identification}

\titlerunning{Feature Bagging for Steganographer Identification}

%
%
\author{Hanzhou Wu}
\authorrunning{H. Wu}

\institute{\email{h.wu.phd@ieee.org}}
%
%

\maketitle

\begin{abstract}
Traditional steganalysis algorithms focus on detecting the existence of steganography in a single object. In practice, one may face a complex scenario where one or some of multiple users also called \emph{actors} are guilty of using steganography, which is defined as the \emph{steganographer identification problem} (SIP). This requires steganalysis experts to design effective and robust detection algorithms to identify the guilty actor(s). The mainstream works use clustering, ensemble and anomaly detection, where distances in high dimensional space between features of actors are determined to find out the outlier(s) corresponding to steganographer(s). However, in high dimensional space, feature points could be sparse such that distances between feature points may become relatively similar to each other, which cannot benefit the detection. Moreover, it is well-known in machine learning that combining techniques such as boosting and bagging can be effective in improving detection performance. This motivates the authors in this paper to present a feature bagging approach to SIP. The proposed work merges results from multiple detection sub-models, each of which feature space is randomly sampled from the raw full dimensional space. We create a new dataset called \emph{ImgNetEase} including 5108 images downloaded from a social website to mimic the real-world scenario. We extract PEV-274 features from images, and take nsF5 as the steganographic algorithm for evaluation. Experiments have shown that our work improves the detection accuracy significantly on created dataset in most cases, which has shown the superiority and applicability.
\end{abstract}

\section{Introduction}
Steganalysis aims to reveal the use of steganography in seemingly-normal objects. Traditional steganalysis algorithms mainly focus on detecting the existence of steganography in a single object. This is treated as a binary classification problem, which motivates people to design effective feature extractors \cite{ShiJPEGMarkov2006}, \cite{PEV274}, \cite{SPAM2010}, \cite{SRM2012} and use supervised classifiers such as SVM. Recently, in-depth studies \cite{XuNet2016}, \cite{XuNetEnsemble2016} are performed on moving deep learning to steganalysis.

In practice, we may face a complex scenario that multiple network actors send a set of media files while one or some of them are using steganography, which is defined as \emph{steganographer identification problem} (SIP) first pointed by Ker \emph{et al.} \cite{Ker2011}. One might use traditional steganalysis algorithms to find stego objects out and then identify the guilty actor(s). However, the guilty actor(s) may be lost due to a number of false positives because of the between-object difference \cite{Ker2012}. It requires us to propose efficient \emph{pooled steganalysis} \cite{batch2006} algorithms.

Though the study of SIP is in its infancy, there have been some works reported in the literature. Currently, the state-of-the-arts \cite{Ker2011}, \cite{Ker2012}, \cite{Li2016}, \cite{Ker2014} mainly use traditional steganalytical features. In their solutions, each actor is represented by a set of feature vectors. The distances between different feature sets corresponding to different actors are determined to measure their similarity. By using hierarchical clustering or local outlier detection, one can collect the suspicious actor(s) that will be judged as the steganographer(s). These methods compute distances in the full dimensional space, which, however, may not always perform well since points in high dimensional space could be sparse, implying that, distances between feature points may become similar to each other, making many normal actors be selected as the guilty ones.

It is true that statistical ensemble of multiple learning algorithms can achieve better prediction performance. To tackle with the aforementioned dimensional problem, we introduce a feature bagging approach in this paper. The proposed approach builds multiple detection sub-models, each of which feature space is sampled from the original full dimensional space. By merging results from sub-models, the most suspicious actor(s) are judged as the steganographer(s). Experiments show that the proposed work can improve the accuracy of detection, which demonstrates superiority and applicability.

The rest of this paper are organized as follows. In Section 2, we introduce the proposed approach. Then, we conduct experiments and analysis for performance evaluation in Section 3. Finally, we conclude this paper in Section 4.

\section{Feature Bagging}
We will consider images as objects. Mathematically, let $A=\{a_1,a_2,...,a_n\},n\geq 2,$ and $S(a_i)=\{\textbf{I}_1^{(i)},\textbf{I}_2^{(i)},...,\textbf{I}_m^{(i)}\},m\geq 1,$ respectively represent actors and the images held by actor $a_i$. A detector computes the \emph{preprocessed} feature vectors for each $a_i$, i.e., $F(a_i)=\{\textbf{f}_1^{(i)},\textbf{f}_2^{(i)},...,\textbf{f}_m^{(i)}\}$. All $F(a_i)~(1\leq i\leq n)$ are divided to disjoint sets with an identical size, i.e.,
\begin{equation}
P(a_i)=\bigcup_{j=1}^{p}{P_j(a_i)},(1\leq i\leq n),
\end{equation}
where $m=p\cdot q$ and $P_j(a_i)=\{\textbf{f}_{jq-q+1}^{(i)},\textbf{f}_{jq-q+2}^{(i)},...,\textbf{f}_{jq}^{(i)}\}$.

Accordingly, each $a_i~(1\leq i\leq n)$ can be represented by $p$ set of feature vectors. We call the $p$ sets as ``$p$ points''. Thus, we can collect a total of $p\cdot n$ points, each of which belongs to one of the $n$ actors. It is naturally assumed that, distances between an abnormal point and a normal point should be larger than that between two normal points. In other words, normal points are densely distributed while abnormal ones are sparsely distributed. Thus, we can utilize anomaly detection for identification, but requiring a well-designed distance measure.

The maximum mean discrepancy (MMD) \cite{MMD2007} has been empirically shown to be quite effective for distance measurement. Given observations $X=\{\textbf{x}_i\}_{i=1}^{|X|}$ and $Y=\{\textbf{y}_i\}_{i=1}^{|Y|}$, which are i.i.d. drawn from $p(\textbf{x})$ and $q(\textbf{y})$ defined on $\mathbb{R}^d$, let $\mathcal{F}$ be a class of functions $f:\mathbb{R}^d\mapsto \mathbb{R}$, the MMD and its empirical estimate are:
\begin{equation}
\text{MMD}[\mathcal{F},p,q] = \underset{f\in \mathcal{F}}{\text{sup}}~\mathbb{E}_{\textbf{x}\sim p(\textbf{x})}f(\textbf{x})-\mathbb{E}_{\textbf{y}\sim q(\textbf{y})}f(\textbf{y}),
\end{equation}
\begin{equation}
\text{MMD}[\mathcal{F},X,Y] = \underset{f\in \mathcal{F}}{\text{sup}}~\frac{1}{|X|}\sum_{\textbf{x}\in X}f(\textbf{x})-\frac{1}{|Y|}\sum_{\textbf{y}\in Y}f(\textbf{y}).
\end{equation}

Usually, $\mathcal{F}$ is selected as a unit ball in a universal RKHS $\mathcal{H}$ defined on compact metric space $\mathbb{R}^d$ with kernel $k(\cdot,\cdot)$ and feature mapping $\phi(\cdot)$. The Gaussian and Laplacian kernels are universal. It is proven that (see Lemma 4 in  \cite{MMD2012}),
\begin{equation}
\text{MMD}^2[\mathcal{F},p,q] = \left \|\mathbb{E}_{\textbf{x}\sim p(\textbf{x})}\phi(\textbf{x})-\mathbb{E}_{\textbf{y}\sim q(\textbf{y})}\phi(\textbf{y}) \right \|_{\mathcal{H}}^2.
\end{equation}
An \emph{unbiased} estimate of MMD is:
\begin{equation}
\text{MMD}[\mathcal{F},X,Y] = \left (\frac{1}{|X|^2-|X|}\sum_{i\neq j}h[i,j]\right )^{1/2},
\end{equation}
where $|X| = |Y|$ is assumed and
\begin{equation}
h[i,j] = k(\textbf{x}_i,\textbf{x}_j) + k(\textbf{y}_i,\textbf{y}_j) - k(\textbf{x}_i,\textbf{y}_j) - k(\textbf{x}_j,\textbf{y}_i).
\end{equation}

For any two points, we use the unbiased estimate of MMD to measure their distance, which has been used in prior arts. However, it is noted that, when a point has only one feature vector, we cannot use MMD since its value always equals zero. In this case, one may use Euclidean metric, i.e., $d(\textbf{x},\textbf{y})=||\textbf{x}-\textbf{y}||_{2}$, or other metrics. Therefore, by using anomaly detection, a ranking list for the $pn$ points is determined according to their anomaly scores. We use $pn$ triples $\{(u_i,v_i,w_i)\}_{i=1}^{pn}$ to denote the sorted information, where $u_1\geq u_2\geq ...\geq u_{pn}$ represent the anomaly scores. $v_i$ denotes the corresponding actor and $w_i$ is the point index, namely, we have $P_{w_i}(v_i)\in P(v_i)$. For each actor $a_i$, we can determine a fusion score below:
\begin{equation}
s(a_i) = \sum_{j=1}^{pn}\frac{(pn+1-j)\cdot \delta(v_j,a_i)}{p},(1\leq i\leq n),
\end{equation}
where $\delta(x,y)=1$ if $x=y$, otherwise $\delta(x,y)=0$. By sorting the fusion scores, we can generate the final ranking list, where the actor with the largest score will be the most suspicious.

\begin{algorithm}[!t]
\caption{Single anomaly detection approach for SIP}
\begin{algorithmic}[1]
\renewcommand{\algorithmicrequire}{\textbf{Input:}}
\renewcommand{\algorithmicensure}{\textbf{Output:}}
\REQUIRE $A=\{a_1,...,a_n\}$, $S(a_i)=\{\textbf{I}_1^{(i)},...,\textbf{I}_m^{(i)}\},i\in [1,n], p.$
\ENSURE A ranking list $\textbf{r}$.
\STATE Extract feature vectors and preprocess them
\STATE Generate disjoint feature sets with Eq. (1)
\STATE Apply outlier detection algorithm $\mathcal{A}$ to $np$ points
\STATE Determine $\{(u_i,v_i,w_i)\}_{i=1}^{pn}$ and apply Eq. (7)
\STATE Sort $\{s(a_i)\}_{i=1}^{n}$ and return a ranking list $\textbf{r}$
\end{algorithmic}
\end{algorithm}

\begin{algorithm}[!t]
\caption{Feature bagging approach for SIP}
\begin{algorithmic}[1]
\renewcommand{\algorithmicrequire}{\textbf{Input:}}
\renewcommand{\algorithmicensure}{\textbf{Output:}}
\REQUIRE $A=\{a_1,...,a_n\}$, $S(a_i)=\{\textbf{I}_1^{(i)},...,\textbf{I}_m^{(i)}\},i\in [1,n],p.$
\ENSURE A ranking list $\textbf{r}_\text{F}$.
\STATE Extract feature vectors and preprocess them
\STATE Generate disjoint feature sets with Eq. (1)
\FOR{$i = 1 \to T$}
    \STATE Produce feature sets with dimension $d_i\in [H/2,H-1]$
    \STATE Apply outlier detection algorithm $\mathcal{A}_i$ to $pn$ ``new'' points
    \STATE Determine $\{(u_i,v_i,w_i)\}_{i=1}^{pn}$ and apply Eq. (7)
    \STATE Sort $\{s(a_i)\}_{i=1}^{n}$ and collect a ranking list $\textbf{r}_i$
\ENDFOR
\STATE Determine the final fusion scores with Eq. (8)
\STATE Sort $\{s_\text{F}(a_1),...,s_\text{F}(a_n)\}$ and return a final ranking list $\textbf{r}_\text{F}$
\end{algorithmic}
\end{algorithm}

In this way, we can construct a single anomaly detection system operated on the full feature space, as shown in \textbf{Algorithm 1}. Feature preprocessing is necessary to guarantee accuracy \cite{Ker2014}. Feature normalization is a good choice and other preprocessing methods may be suitable as well such as principal component transformation. By normalization, each feature component has zero mean and unit variance. The preprocessing enables the distance measure to be more meaningful and not significantly affected by noisy components.

A kernel function is required when to use MMD. Ker \emph{et al.} \cite{Ker2014} have examined multiple kernels such as linear kernel, Gaussian kernel and the centroid `kernel'. It is believed that, alternative kernels have advantages against certain batch embedding strategies. From the point of computational complexity, both linear kernel and the centroid kernel are desirable. In default, we recommend the linear kernel, i.e., $k(\textbf{x},\textbf{y})=\textbf{x}\cdot\textbf{y}$. It is proved that, the centroid `kernel' approximates the true linear MMD for large size of samples (see Appendix in \cite{Ker2014}).

Due to the sparse nature in the high dimensional space \cite{HDD2001}, the distances among points may become similar which cannot benefit anomaly detection. The proposed approach uses feature bagging to deal with this problem and improve the accuracy of detection, where the dimension of feature vectors is reduced. However,
the number of feature vectors are unchanged because we believe that reducing the number of feature vectors could reduce ``signal-to-noise ratio'', which cannot benefit detection. Actually, we conducted experiments and found that, reducing the number of feature vectors reduces detection performance. However, we admit that, the decrease of accuracy should be also affected by the image diversity and other potential factors.

Mathematically, the proposed work will build $T$ sub-models $\mathcal{M}=\{\mathcal{M}_1,\mathcal{M}_2,$ ..., $\mathcal{M}_T\}$, whose feature dimension vector is denoted by $\textbf{d}=(d_1,d_2,...,d_T)$. Each $d_i~(1\leq i\leq T),$ is chosen from the range $[H/2,H-1]$, where $H$ is the dimension of the raw full feature space. It is possible that $d_i=d_j$ for some $i\neq j$. Each $\mathcal{M}_i~(1\leq i\leq T)$ corresponds to a single anomaly detection system similar to \textbf{Algorithm 1}, where a ranking list $\textbf{r}_i$ can be collected. The only difference is that, $\mathcal{M}_i$ uses the $d_i$-D \emph{random} subspace of the original $H$-D space. By further processing $\{\textbf{r}_1,\textbf{r}_2,...,\textbf{r}_T\}$, the final fusion score for each $a_i~(1\leq i\leq n)$ can be generated as follows:
\begin{equation}
s_\text{F}(a_i) = \sum_{j=1}^{T}\frac{n+1-\sum_{k=1}^{n}[k\cdot \delta(r_k^{(j)},a_i)]}{T},
\end{equation}
where $\textbf{r}_j$ = $(r_1^{(j)}$, ..., $r_n^{(j)})$, and $r_k^{(j)}$ means the actor with the $k$-th largest anomaly score. Namely, for $\textbf{r}_j$, $r_1^{(j)}$ is the most suspicious and $r_n^{(j)}$ is the least suspicious. By sorting $\{s_\text{F}(a_1), ..., s_\text{F}(a_n)\}$, we can generate the final ranking list, where the actor with the largest score will be the most suspicious, and the smallest score corresponds to the least suspicious. \textbf{Algorithm 2} shows the procedure.

\begin{figure}[!t]
\centering
\includegraphics[width=4.8in]{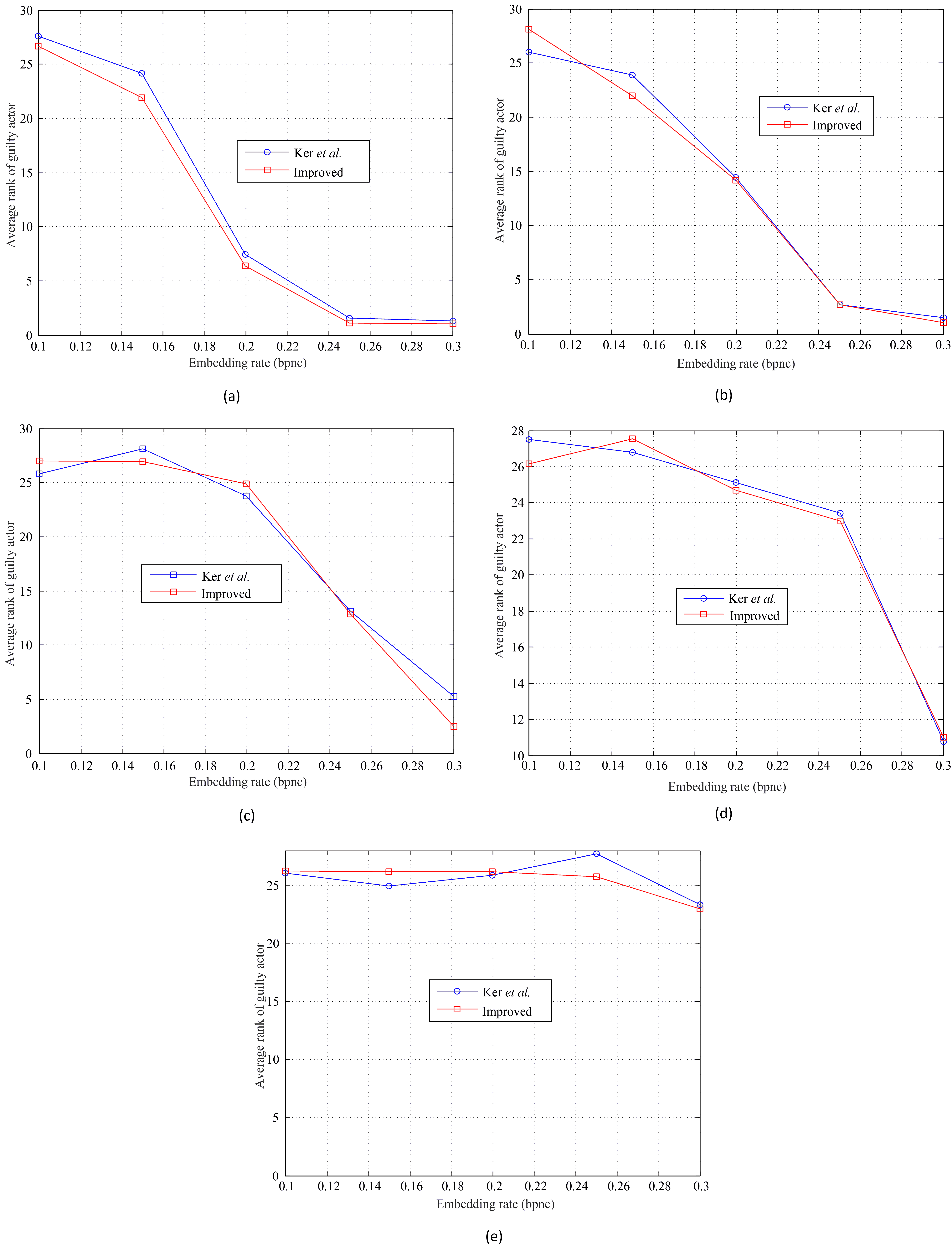}
\caption{Performance comparison between Ker \emph{et al.}'s method and its improved version using feature bagging with different QFs:
(a) QF = 70, (b) QF = 75, (c) QF = 80, (d) QF = 85 and (e) QF = 90.
}
\end{figure}

\section{Performance Evaluation}
In this section, we will conduct experiments for evaluation.

\textbf{Database:} We take JPEG images as the objects held by actors. To mimic the real-world scenario, we use a web crawler to download images from a Chinese open-leading social network site \emph{NetEase}\footnote{https://www.163.com/}. The resultant \emph{ImgNetEase} database contains 5108 images, among which around 90\% are with a quality factor (QF) close to 90. The average size is close to $1120~(\text{height})\times 1600~(\text{width})$. The images are very diverse.

\textbf{Embedding Algorithm:} Prior arts \cite{Li2016}, \cite{Ker2014} take F5, nsF5, JPHide\&Seek, OutGuess and StegHide as the steganographic algorithms to be tested. We here sincerely refer a reader to \cite{Li2016}, \cite{Ker2014} for the brief introduction of these steganographic algorithms. It has been shown that, among these algorithms, nsF5 is the most secure. For simplicity, we take nsF5 as the embedding tool in this paper.

\textbf{Steganalysis Features:} In our experiments, each image will be represented by a 274-D feature vector called PEV-274 \cite{PEV274}, designed for JPEG steganalysis and previously shown to be effective against nsF5. PEV-274 was used in \cite{Ker2011}, \cite{Ker2012}, \cite{Ker2014}. For fair comparison, we take PEV-274 as the feature extractor. Notice that, nsF5 is detectable by modern steganalysis features.

\textbf{Embedding Strategy:} A guilty actor should choose how to divide a message to multiple pieces, each of which is carried by a selected image.
It poses new optimization problem, which, however, is not the main interest of this paper. For simplicity, we choose the even strategy \cite{Ker2012Strategy}, i.e., all cover images will carry the same payload regardless of their secure capacity. Notice that, there has no strategy that is proven to be theoretically optimal.

\textbf{Outlier Detection:} The local outlier factor (LOF) \cite{LOF2000} will be used for anomaly detection. In LOF, unless mentioned, an integer $k$ specifying the number of nearest neighbors of a point is set as 10. The MMD is chosen as the distance measure in case $p\neq m$, and Euclidean distance for $p=m$. We preprocess feature vectors by normalization, and use linear kernel for MMD.

We use $T=16$ sub-models for feature bagging. All raw images in \emph{ImgNetEase} are cropped from their central regions to create 5 new image datasets, where images are sized $512\times 512$ and the QFs are 70, 75, 80, 85 and 90, respectively. It simplifies steganalysis since steganalysis features are sensitive to different quantisation matrices \cite{Ker2014}, \cite{multiclass}. We denote the datasets by SetCover-70, SetCover-75, SetCover-80, SetCover-85, and SetCover-90. For each dataset, we apply the nsF5 simulator\footnote{http://dde.binghamton.edu/download/nsf5simulator/} with 5 data-embedding rates, resulting in 5 stego datasets. The data-embedding rates are 0.1, 0.15, 0.2, 0.25, 0.3 bits per non-zero coefficient (bpnc), e.g., for SetCover-70, we generate 5 datasets, denoted by SetStego-70-0.1, SetStego-70-0.15, SetStego-70-0.2, SetStego-70-0.25, and SetStego-70-0.3.

In each experiment, we take $n=50$ and $m=100$. Exactly one guilty actor is simulated by using nsF5. For each combination of parameters, each experiment is repeated 100 times with a random selection of the index number of the guilty actor. We use the average rank of the guilty actor as the metric to reflect how well the guilty actor is identified. We take SetCover-70 and SetStego-70-0.1 for better explanation. We randomly choose 5000 images images from SetCover-70, and randomly divide them to 50 groups, each of which belongs to an actor $a_i~(1\leq i\leq 50)$. Then, we randomly generate an index $1\leq g\leq 50$, and replace the cover images held by $a_g$ with the corresponding stego images in SetStego-70-0.1. By extracting steganalysis features and applying outlier detection, we rank all actors according to their anomaly scores. Afterward, by repeating the process with 100 times, we can determine the average rank of the guilty actor.

Fig. 1 demonstrates the performance comparison between Ker \emph{et al.}'s method \cite{Ker2012}, \cite{Ker2014} and its improved version using feature bagging with different QFs. We simulate Ker \emph{et al.}'s method \cite{Ker2012}, \cite{Ker2014} with the above-mentioned configurations. We use $p=1$ as we found $1< p< m$ cannot achieve better improvement when feature bagging is applied, which may be due to multiple factors such as feature sensitivity, image diversity and reduction of `signal-to-noise ratio'.

It is observed from Fig. 1 that, better detection performance can be achieved with a smaller QF no matter feature bagging is applied or not. This follows the empirical result in steganalysis. When QF = 90, the average ranks of the guilty due to different embedding rates are all close to 25, which corresponds to random guessing. Moreover, for different QFs, when the embedding rate is relatively low (e.g., 0.1 bpnc), the detection performance also corresponds to random guessing. It has indicated the difficulty of steganalysis at low data embedding rates. It can be also seen that, with feature bagging, the detection performance can be further improved in most cases, which has shown the potential of feature bagging. Indeed, we can find that, in some cases, feature bagging provides worse performance, which is normal as we did not optimize the feature selection.

Fig. 1 (e) has shown that Ker \emph{et al.}'s method and the method using feature bagging are equivalent to random guessing with the corresponding parameters. One might think that it is mainly due to the steganalysis features (PEV-274). However, we point that, it could be majorly due to the MMD distance since we find that, surprisingly, replacing the MMD distance with Euclidean distance (where $p=m$ is required) results in effective detection performance, which can be observed from Fig. 2. As shown in Fig. 2, feature bagging still has the potential to improve the performance. And, though the Euclidean distance does not outperform the MMD distance in case QF = 70, the former provides effective and better performance in case QF = 90. It indicates that, regardless of the steganalysis features, a well-designed distance measure is required for achieving superior detection performance, which should be a core topic for SIP.

\begin{figure}[!t]
\centering
\includegraphics[width=4in]{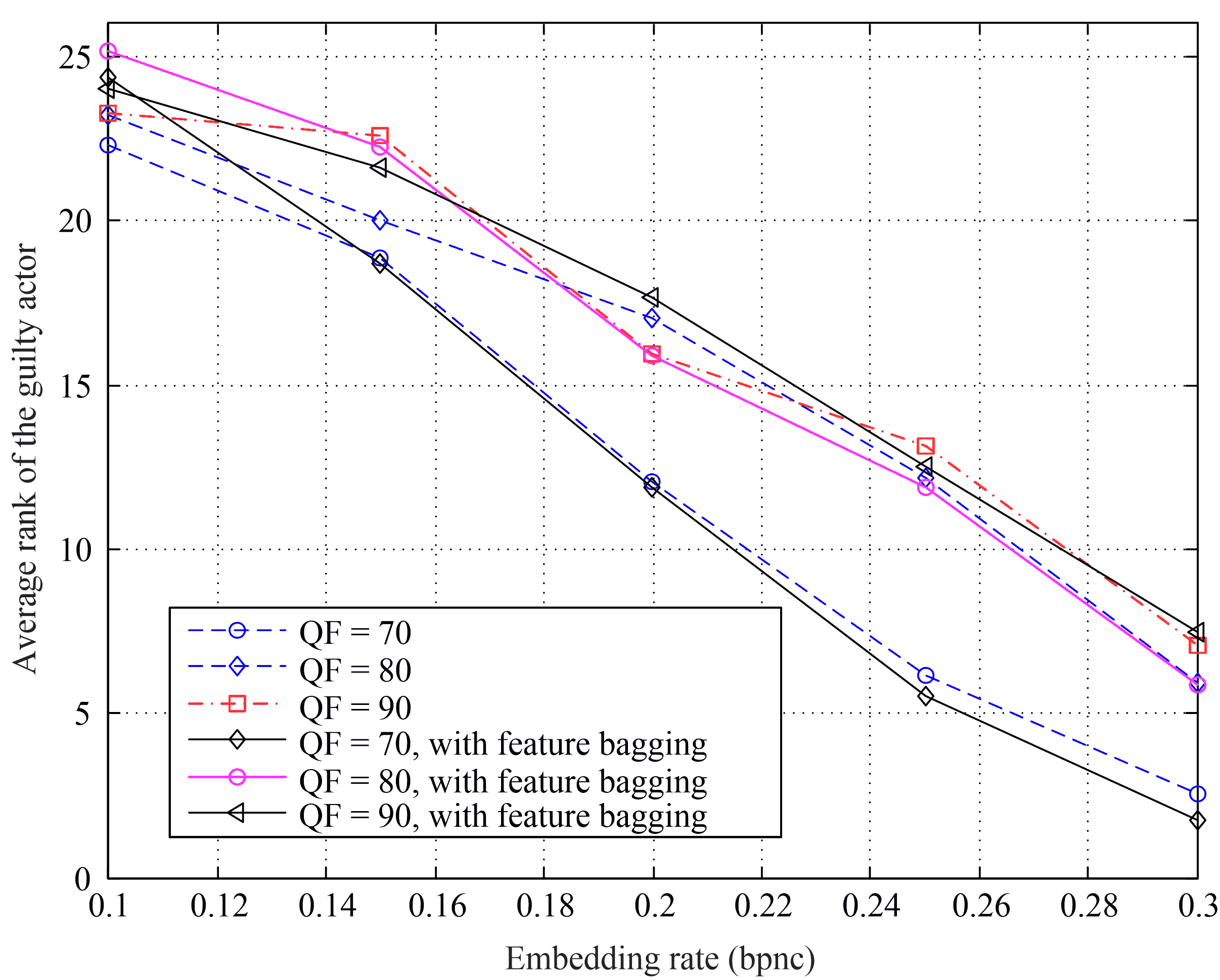}
\caption{Detection performance using the Euclidean distance ($p=m$).}
\end{figure}

\section{Conclusion}
In this paper, we present a simple but effective feature bagging approach for the SIP. The proposed work combines the detection results from sub-models, each of which feature space is randomly sampled from the raw full dimensional space. Experimental results show that our work has the ability to improve the detection performance in most cases, which has demonstrated the superiority of our work. Form the viewpoint of performance optimization, there is still room for improvement. For example, one may design specific feature selection algorithm, rather than random selection, for choosing efficient feature components for detection. Designing effective steganalysis features is also quite necessary. Our future focus will be the steganalysis features, and the design of distance measure between feature sets.

\end{document}